\documentstyle[12pt]{article}
\newlength{\extraspace}
\newlength{\extraspaces}
\newcommand{\be}{\begin{equation}
\addtolength{\abovedisplayskip}{\extraspaces}
\addtolength{\belowdisplayskip}{\extraspaces}
\addtolength{\abovedisplayshortskip}{\extraspace}
\addtolength{\belowdisplayshortskip}{\extraspace}}
\newcommand{\ee}{\end{equation}}

\newcommand{\ba}{\begin{eqnarray}
\addtolength{\abovedisplayskip}{\extraspaces}
\addtolength{\belowdisplayskip}{\extraspaces}
\addtolength{\abovedisplayshortskip}{\extraspace}
\addtolength{\belowdisplayshortskip}{\extraspace}}
\newcommand{\ea}{\end{eqnarray}}

\newcommand{\nonu}{\nonumber \\[.5mm]}
\newcommand{\A}{&\!\!\!}

\newcommand{\il}{\lambda_{{}_{{}_{\!\!\!\!\scriptstyle{i}}}}}

\newcommand{\bl}{\lambda_{{}_{{}_{\!\!\!\!\scriptstyle{0}}}}}

\newcommand{\newsection}[1]{
\vspace{7mm} \pagebreak[3] \addtocounter{section}{1}
\setcounter{subsection}{0} \setcounter{footnote}{0}
\begin{center}
{\large {\bf \thesection. #1}}
\end{center}
\nopagebreak
\medskip
\nopagebreak \hspace{3mm}}

\begin{document}
\begin{center}
{{\bf Vacuum Non Singular Black Hole Solutions in Tetrad Theory of
Gravitation}}
\end{center}
\centerline{ Gamal G.L. Nashed}

\bigskip

\centerline{{\it Mathematics Department, Faculty of Science, Ain
Shams University, Cairo, Egypt }}

\bigskip
 \centerline{ e-mail:nasshed@asunet.shams.eun.eg}

\hspace{2cm}
\\
\\
\\
\\
\\
\\
\\
\\
Starting from a spherically symmetric tetrad with three unknown
functions of the radial coordinate, a general solution of M\o
ller's field equations in case of spherical symmetry nonsingular
black hole is derived. The previously obtained solutions are
verified as special cases of the general solution. The general
solution is characterized by an arbitrary function and two
constants of integration. The general solution gives no more than
the spherically symmetric nonsingular black hole solution. The
energy content of the general solution depends on the asymptotic
behavior of the arbitrary function, and is different from the
standard one.

\newpage
\begin{center}
\newsection{\bf Introduction}
\end{center}

M\o ller has shown that the problem of the energy-momentum complex
has no solution in the framework of gravitational field theories
based on Riemannian space \cite{Mo}. In a series of papers,
\cite{Mo,Mo3,Mo4} he was able to obtain
 a general expression for a satisfactory energy-momentum complex in the absolute
 parallelism space. The Lagrangian formulation
of the theory was given by Pellegrini and Plebanski \cite{PP}.
Quite independently Hayashi and Nakano \cite{HN} formulated the
tetrad theory of gravitation as a gauge theory of the space-time
translation group. In these attempts, the admissible Lagrangians
were limited by the assumption that the field equations has the
Schwarzschild solution. M\o ller \cite{Mo1} abandoning this
assumption and look for a wider class of Lagrangians by
constructing a new field theory. His aim was to get a theory free
from singularity while retaining the main merits of general
relativity as far as possible. Meyer \cite{Me} showed that M\o
ller's theory is a special case of the Poincar$\acute{e}$ gauge
theory \cite{HS,HNV}.  S$\acute{a}$ez \cite{Se} generalized M\o
ller theory into a scalar tetradic theory of gravitation.

In an earlier paper \cite{Ga} the author used a spherically
symmetric tetrad constructed by Robertson \cite{Ro} to derive two
different spherically symmetric vacuum nonsingular black hole
solutions of M\o ller's field equations. He also calculated the
energy content of these solutions \cite{Ga}. It is the purpose of
the present work to derive the general solution of M\o ller's
tetrad theory of gravitation assuming a specific form of the
stress-energy momentum tensor as given by Dymnikova \cite{Di},
then calculated the energy content of this general solution.

In section 2 we briefly review M\o ller's tetrad theory of
gravitation. In section 3 the structure of the tetrad spaces
having spherical symmetry as well as the previously obtained
solutions of M\o ller's field equations are reviewed. Assuming a
specific form of the stress-energy momentum tensor, the general
solution of M\o ller field equations is derived in section 4. The
energy content of this general solution is derived in section 5.
Section 6 is devoted to the main results and conclusions.

Computer algebra system Maple V Release 4 is used in some
calculations.

\newsection{M\o ller's  tetrad theory of gravitation}

M\o ller's constructed a gravitational theory based on
 Weitzenb{\rm $\ddot{o}$}ck space-time. In this
theory the field variables are the 16 tetrad components $\il^\mu$,
from which the metric is derived by \be g^{\mu \nu} \stackrel{\rm
def.}{=} \il^\mu \il^\nu. \ee
 We assume an imaginary values for the vector $\bl^\mu$ in order to have a Lorentz signature.
 We note that, associated with any tetrad field $\il^\mu$ there is a metric field defined
 uniquely by (1), while a given metric $g^{\mu \nu}$ does not
 determine the tetrad field completely; for any local Lorentz
 transformation of the tetrads $\il^\mu$ leads to a new set of
 tetrads which also satisfy (1).
  The Lagrangian ${\it L}$ is an invariant constructed from
$\gamma_{\mu \nu \rho}$ and $g^{\mu \nu}$, where $\gamma_{\mu \nu
\rho}$ is the contorsion tensor given by \be \gamma_{\mu \nu \rho}
\stackrel{\rm def.}{=} {\il}_{\,{}^{{}^{{}^{\scriptstyle{\mu}}}}}
{\il}_{\,{}^{{}^{{}^{\scriptstyle{\nu;\rho}}}}}, \ee where the
semicolon denotes covariant differentiation with respect to
Christoffel symbols. The most general Lagrangian density invariant
under the parity operation is given by the form \be {\cal L}
\stackrel{\rm def.}{=} (-g)^{1/2} \left( \alpha_1 \Phi^\mu
\Phi_\mu+ \alpha_2 \gamma^{\mu \nu \rho} \gamma_{\mu \nu \rho}+
\alpha_3 \gamma^{\mu \nu \rho} \gamma_{\rho \nu \mu} \right), \ee
where \be g \stackrel{\rm def.}{=} {\rm det}(g_{\mu \nu}),
 \ee
 and
$\Phi_\mu$ is the basic vector field defined by \be \Phi_\mu
\stackrel{\rm def.}{=} {\gamma^\rho}_{\mu \rho}. \ee Here
$\alpha_1, \alpha_2,$ and $\alpha_3$ are constants determined by
M\o ller
 such that the theory coincides with general relativity in the weak fields:

\be \alpha_1=-{1 \over \kappa}, \qquad \alpha_2={\lambda \over
\kappa}, \qquad \alpha_3={1 \over \kappa}(1-2\lambda), \ee where
$\kappa$ is the Einstein constant and  $\lambda$ is a free
dimensionless parameter\footnote{Throughout this paper we use the
relativistic units, $c=G=1$ and
 $\kappa=8\pi$.}. The same
choice of the parameters was also obtained by Hayashi and Nakano
\cite{HN}.

M\o ller applied the action principle to the Lagrangian density
(3) and obtained the field equation in the form \be G_{\mu \nu}
+H_{\mu \nu} = -{\kappa} T_{\mu \nu}, \ee \be F_{\mu \nu}=0, \ee
where the Einstein tensor $G_{\mu \nu}$ is defined by \be G_{\mu
\nu}=R_{\mu \nu}-{1 \over 2} g_{\mu \nu} R. \ee Here $H_{\mu \nu}$
and $F_{\mu \nu}$ are given by \be H_{\mu \nu} \stackrel{\rm
def.}{=} \lambda \left[ \gamma_{\rho \sigma \mu} {\gamma^{\rho
\sigma}}_\nu+\gamma_{\rho \sigma \mu} {\gamma_\nu}^{\rho
\sigma}+\gamma_{\rho \sigma \nu} {\gamma_\mu}^{\rho \sigma}+g_{\mu
\nu} \left( \gamma_{\rho \sigma \lambda} \gamma^{\lambda \sigma
\rho}-{1 \over 2} \gamma_{\rho \sigma \lambda} \gamma^{\rho \sigma
\lambda} \right) \right],
 \ee
and \be F_{\mu \nu} \stackrel{\rm def.}{=} \lambda \left[
\Phi_{\mu,\nu}-\Phi_{\nu,\mu} -\Phi_\rho \left({\gamma^\rho}_{\mu
\nu}-{\gamma^\rho}_{\nu \mu} \right)+ {{\gamma_{\mu
\nu}}^{\rho}}_{;\rho} \right], \ee and they are symmetric and skew
symmetric tensors, respectively.

M\o ller assumed that the energy-momentum tensor of matter fields
is symmetric. In the Hayashi-Nakano theory, however, the
energy-momentum tensor of spin-$1/2$ fundamental particles has
nonvanishing antisymmetric part arising from the effects due to
intrinsic spin, and the right-hand side of (8) does not vanish
when we take into account the possible effects of intrinsic spin.

It can be shown \cite{HS1} that the tensors, $H_{\mu \nu}$ and
 $F_{\mu \nu}$, consist of only those terms which are linear or quadratic
in the axial-vector part of the torsion tensor, $a_\mu$, defined
by \be a_\mu \stackrel{\rm def.}{=} {1 \over 3} \epsilon_{\mu \nu
\rho \sigma} \gamma^{\nu \rho \sigma}, \ee where $\epsilon_{\mu
\nu \rho \sigma}$ is defined by \be \epsilon_{\mu \nu \rho \sigma}
\stackrel{\rm def.}{=} (-g)^{1/2} \delta_{\mu \nu \rho \sigma} \ee
with $\delta_{\mu \nu \rho \sigma}$ being completely antisymmetric
and normalized as $\delta_{0123}=-1$. Therefore, both $H_{\mu
\nu}$ and $F_{\mu \nu}$ vanish if the $a_\mu$ is vanishing. In
other words, when the $a_\mu$ is found to vanish from the
antisymmetric part of the field equations, (8), the symmetric part
(7) coincides with the Einstein equation.

\newsection{Spherically symmetric nonsingular black hole solutions}
The structure of the Weintzenb${\ddot o}$ck spaces with spherical
symmetry has been studied by Robertson \cite{Ro}. The tetrad space
having three unknown functions of radial coordinate with spherical
symmetry in spherical polar coordinates, can be written as
 \be
\left(\il^\mu \right)= \left( \matrix{ iA & iDr & 0 & 0
\vspace{3mm} \cr 0 & B \sin\theta \cos\phi & \displaystyle{B \over
r}\cos\theta \cos\phi
 & -\displaystyle{B \sin\phi \over r \sin\theta} \vspace{3mm} \cr
0 & B \sin\theta \sin\phi & \displaystyle{B \over r}\cos\theta
\sin\phi
 & \displaystyle{B \cos\phi \over r \sin\theta} \vspace{3mm} \cr
0 & B \cos\theta & -\displaystyle{B \over r}\sin\theta  & 0 \cr }
\right), \ee where the vector $\bl^\mu$ has taken to be imaginary
in order to preserve the Lorentz signature for the metric, i.e,
the functions $A$ and $D$ have to be taken as imaginary.

Applying (14) to the field equations (7) and (8) we note that the
two tensors $H_{\mu \nu}$ and  $F_{\mu \nu}$ are vanishing
identically regardless of the values of the functions $A$, $B$ and
$D$. Thus M\o ller field equations reduce for the tetrad (14) to
Einstein's equations (9). The corresponding  field equations (7)
and (8) have given rise to the following equations.
\newpage
\ba \kappa T_{0 0} \A= \A {1 \over r A^2 B^4}\Biggl[ \Biggl \{
\Biggl(3 D^2+8B'^2\Biggr) D-2\Biggl(2D B''+B'D'\Biggr)B \Biggr \}
r^3 B^2D- \nonu
\A \A \Biggl \{2\Biggl(D B''+B'D'\Biggr)B-5DB'^2\Biggr \} r^5
D^3-\Biggl(2BB''-3D^2-3B'^2 \Biggr) rB^4+\nonu
\A \A 2\Biggl(BD'-4DB'\Biggr)
r^4BD^3+2\Biggl(BD'-6DB'\Biggr)r^2B^3D-4B^5B' \Biggr], \nonu
\kappa T_{0 1} \A= \A  {D \over  A B^4} \Biggl[ \Biggl \{ 2 \Biggl
(D B''+B'D'\Biggr) B-5DB'^2  \Biggr \}
r^3D+\Biggl(2BB''-3D^2-3B'^2 \Biggr) rB^2-\nonu
\A \A  2\Biggl(B D'-4DB'\Biggr)r^2BD+4B^3B' \Biggr], \nonu
\kappa T_{1 1} \A= \A {1 \over r A B^4}\Biggl[ \Biggl \{ \Biggl(3
D^2+B'^2\Biggr) A+2BA'B' \Biggr \} r B^2- \Biggl \{ 2 \Biggl (D
B''+B'D'\Biggr) B-5DB'^2  \Biggr \} r^3AD + \nonu
\A \A  2\Biggl(B D'-4DB'\Biggr)r^2ABD-2AB^3B'-2B^4A' \Biggr],\nonu
\kappa T_{2 2} \A= \A {r \over  A^2 B^4} \Biggl[ \Biggl ( \Biggl
\{ \Biggl(D A''+3 A'D'\Biggr) B-3DA'B'  \Biggr \} A B D +\Biggl\{
\Biggl (2D B''+5B'D'\Biggr) B D- \nonu
\A \A \Biggl( D D''+D'^2 \Biggr) B^2-5D^2B'^2\Biggr \}
A^2-2B^2D^2A'^2 \Biggr) r^3+\nonu
\A \A \Biggl \{ \Biggl(B'^2 -3D^2\Biggr)
A^2-AB^2A''-B''BA^2+2B^2A'^2 \Biggr\}rB^2-\nonu
\A \A 2\Biggl \{ \Biggl( 3BD'-4DB'\Biggr)A-2BDA' \Biggr \}
r^2ABD+A^2B^3B'+AB^4A' \Biggr ], \nonu
T_{3 3} \A= \A sin \theta^2 T_{2 2}, \ea where
$A'=\displaystyle{dA \over dr}$, $B'=\displaystyle{dB \over dr}$
and  $D'=\displaystyle{dD \over dr}$.

Now we are going to review some exact solutions to the partial
differential equations (15). A first trivial flat space-time
solution for the field equations (15) is obtained by taking \be
A=1, \qquad B=1, \qquad D=0. \ee A second non-trivial solution can
be obtained by taking  $A=1$, $B=1$, $D\neq 0$ and solving for
$D$, the result is \be A=1, \qquad B=1 \qquad D=\sqrt{{2m\over
r3}\left(1-e^{-r^3/{r_1}^3}\right)}. \ee A third non-trivial
solution can be obtained by taking $D=0$ and solve for A and B.
 This case is
studied by the author \cite{Ga} where he obtained
\newpage
\ba A \A=\A {1 \over \sqrt{1- \displaystyle{2m \over
R}\left(1-e^{-R^3/{r_1}^3}\right)}}, \nonu
B \A=\A \sqrt{1-\displaystyle{2m \over
R}\left(1-e^{-R^3/{r_1}^3}\right)}. \ea where the stress-energy
momentum tensor has the form \cite{Ga} \ba
{T_0}^0={T_1}^1 \A=\A \epsilon_0 e^{-R^3/{r_1}^3},\nonu %
{T_2}^2={T_3}^3 \A=\A \epsilon_0 e^{-R^3/{r_1}^3} \left(1-{3R^3
\over 2{r_1}^3} \right),
 \ea
 here $R$ and $r_1$ are defined as
 \ba
 R \A= \A {r \over B} \nonu
 {r_1}^3 \A=\A r_g {r_0}^2,\nonu
 r_g\A=\A 2m,\nonu
{r_0}^2 \A=\A {3 \over 8\pi\epsilon_0}.
 \ea

\newsection{General Black Hole solution of M\o ller's Field Equations}
Mikhail and Wanas \cite{MW} constructed a generalized field theory
based on the  Weintzenb${\ddot o}$ck space. Wanas \cite{Wa}
obtained a spherically symmetric solutions using the tetrad (14)
in the case of +ve definite. Mazumder and Ray \cite{MR} completely
integrated the field equations of Mikhail and Wanas for the tetrad
(14) by a suitable change of variables. Mikhail et al. \cite{MWLH}
obtained a general solution in M\o ller's tetrad theory of
gravitation for the tetrad (14)
 in the vacuum case. It is our purpose to find a general solution for
 the tetrad (14) when the stress-energy momentum tensor is not vanishing
 and has the form
\ba
 {T_0}^0={T_1}^1, \nonu
  {T_2}^2={T_3}^3,
 \ea
  where all the other mixed spatial components equal to zero
  \cite{Di}. Then the left hand side of the second equation of
  equations (15) is equal zero and we can find a solution of the
  unknown function $D$ in terms of the unknown function $B$
  in the form
\be D=\displaystyle{1 \over \left(1-\displaystyle{r B' \over B}
\right)}
   \sqrt{ \displaystyle{k_1 B^3 \over r^3}
  \left(1-{\huge e^{(-r^3/{r_1}^3)}} \right)+ \displaystyle{B B' \over r}
  \left(\displaystyle{r B' \over B}  -2 \right)},
\ee where $k_1$ is a constant of integration. From the first and
third equations of (15) using (21) and (22), we get the unknown
function $A$ in the form \be A=\displaystyle{k_2 \over
\left(1-\displaystyle{r B' \over B} \right)}, \ee with $k_2$ being
another  constant of integration. The general solution (22) and
(23) satisfy the field equations (15) when the stress-energy
momentum tensor has the form \ba {T_0}^0={T_1}^1 \A=\A
\displaystyle{\epsilon_0 k_1 B^3 e^{-r^3/{r_1}^3} \over  2m
\left(1-\displaystyle{r B' \over B} \right)},\nonu
{T_2}^2={T_3}^3 \A=\A \displaystyle{\epsilon_0 k_1 B^3
e^{-r^3/{r_1}^3} \left [\left(1-\displaystyle{3r^3 \over
2{r_1}^3}\right)+\displaystyle{3 r^3 \over 2 {r_1}^3} \left(
\displaystyle{r B' \over B}\right)+\displaystyle{r^2 B'' \over 2
B}
 -\left(\displaystyle{r B' \over B}\right)^2 \right] \over 2m\left(1-\displaystyle{r B' \over B} \right)^3},
\ea

The line-element squared of (14) takes the form \be
ds^2=-{(B^2-D^2 r^2) \over A^2B^2} dt^2-{2Dr \over AB^2}dr dt+ {1
\over B^2} (dr^2+r^2 d\Omega^2) \ee with
${d\Omega^2=d\theta^2+\sin^2\theta d\phi^2}$. We assume $B(r)$ to
be nonvanishing so that the surface area of the sphere of a
constant ${\it r}$ be finite. We also assume that $A(r)$ and
$B(r)$ satisfy the asymptotic condition,
 $\lim_{r \to \infty} A(r)$=$\lim_{r \to \infty} B(r)=1$ and
 $\lim_{r \to \infty} rB'=0$. Then, we can show from  (22), (23) and (25)
 that\\
(1) $k_2=1$,\\ (2)\, $B(r)>0$,\\ (3) \, $\lim_{r \to \infty}
rD(r)=0,$ \quad
and \\
(4) if $B-r B'$ vanishes at some point, then
$\left(1-\displaystyle{ B(r) k_1 \left(1-{\huge e^{(-r^3/r1^3)}}
\right) \over r }\right)<0 $ at that point.

Using the coordinate transformation \\
\be dT=dt+{ADr \over B^2-D^2r^2}dr, \ee we can eliminate the cross
term of (25) to obtain \ba ds^2= -\eta_1 dT^2 +{1 \over \eta_1}
{dr^2 \over A^2B^2} + {r^2 \over B^2}d\Omega^2 \ea with
$\eta_1={(B^2-D^2 r^2)/A^2B^2}$. Taking the new radial coordinate
$R={r/B}$, we finally get \be ds^2= -\eta_1 dT^2 +{dR^2 \over
\eta_1} +R^2d\Omega^2, \ee where \be \eta_1(R)=\left(
1-{k_1\left(1-{\huge e^{(-R^3/{r_1}^3)}} \right) \over R} \right).
\ee Then, (29) coincides with the nonsingular black hole solution
given before by Dymnikova \cite{Di} with the mass, $m= {k_1/2}$,
and hence the general solution in the case of the spherically
symmetric tetrad when the stress-energy momentum tensor is
nonvanishing gives no more than the nonsingular black hole
solution  when $1-{rB'/B}$ has no zero and ${\it R}$ is
monotonically increasing function of ${\it r}$. If $1-{rB'/B}$ has
zeroes, the line-element (25) is singular at these zeroes which
lie inside the event horizon as is seen from the property (4)
mentioned above. We shall study in the future whether this
singularity at zero-points of $1-{rB'/B}$ is physically acceptable
or not.

 After using the above transformations, the tetrad (14) can be put in the form
 \be
\left(\il^\mu \right)= \left( \matrix{ \displaystyle{i{\cal A}
\over 1-{\cal D}^2 R^2} & i{\cal D} R(1-R{\cal B}') & 0 & 0
\vspace{3mm} \cr \displaystyle{{\cal A} {\cal D} R  \sin\theta
\cos\phi \over 1-{\cal D}^2R^2} &(1-R {\cal B}') \sin\theta
\cos\phi & \displaystyle{\cos\theta \cos\phi \over R}
 & -\displaystyle{\sin\phi \over R \sin\theta} \vspace{3mm} \cr
\displaystyle{{\cal A} {\cal D} R  \sin\theta \sin\phi  \over
1-{\cal D}^2 R^2} & (1-R{\cal B}') \sin\theta \sin\phi &
 \displaystyle{\cos\theta \sin\phi \over R}
 & \displaystyle{\cos\phi \over R \sin\theta} \vspace{3mm} \cr
\displaystyle{{\cal A} {\cal D} R \cos\theta  \over 1-{\cal
D}^2R^2} & (1-R{\cal B}') \cos\theta & \displaystyle{-\sin\theta
\over R} & 0 \cr } \right), \ee Here ${\cal A}$ and ${\cal D}$ are
given in terms of the unknown function $B(R)$ as \ba
 {\cal A(R)} \A=\A \displaystyle{1 \over 1-R {\cal B}'},\nonu
{\cal D(R)} \A=\A\displaystyle{1 \over 1-R {\cal B}'}
\sqrt{\displaystyle{2m \over R^3} \left(1-{\huge
e^{(-R^3/{r_1}^3)}} \right)+ \displaystyle{{\cal B}' \over R}
\left(R {\cal B}' -2 \right)}, \ea where ${\cal
B}'=\displaystyle{d{\cal B(R)} \over dR}$. It is of interest to
note that the general solution (31) satisfies the field equations
of M\o ller's theory when the stress-energy momentum tensor has
the form \ba {T_0}^0={T_1}^1 \A=\A \epsilon_0
e^{-R^3/{r_1}^3},\nonu
{T_2}^2={T_3}^3 \A=\A \epsilon_0  e^{-R^3/{r_1}^3}
\left(1-\displaystyle{3R^3 \over 2{r_1}^3}\right), \ea

The previously obtained solutions can be verified as  special
cases of the general solution (31). The choice \be {\cal B(R)}=1,
\ee reproduces the solution (17). On the other hand, the choice
\be {\cal B(R)}=\int{{1 \over
R}\left(1-\sqrt{1-\displaystyle{2m(1- e^{-R^3/{r_1}^3}) \over
R}}\right)}dR, \ee
 reproduces the solution
(18). It is of interest to note that if the exponential term is
equal zero then the general solution (31) reduces to that obtained
before by Mikhail et al. \cite{MWLH} and the two choices (33) and
(34) will give the Schwarzschild solution in its standard form.

\newsection{The Energy Associated with the General Solution}

The superpotential of M\o ller's theory is given by Mikhail et al.
\cite{MWHL} as \be {{\cal U}_\mu}^{\nu \lambda} ={(-g)^{1/2} \over
2 \kappa} {P_{\chi \rho \sigma}}^{\tau \nu \lambda}
\left[\Phi^\rho g^{\sigma \chi} g_{\mu \tau}
 -\lambda g_{\tau \mu} \gamma^{\chi \rho \sigma}
-(1-2 \lambda) g_{\tau \mu} \gamma^{\sigma \rho \chi}\right], \ee
where ${P_{\chi \rho \sigma}}^{\tau \nu \lambda}$ is \be {P_{\chi
\rho \sigma}}^{\tau \nu \lambda} \stackrel{\rm def.}{=}
{{\delta}_\chi}^\tau {g_{\rho \sigma}}^{\nu \lambda}+
{{\delta}_\rho}^\tau {g_{\sigma \chi}}^{\nu \lambda}-
{{\delta}_\sigma}^\tau {g_{\chi \rho}}^{\nu \lambda} \ee with
${g_{\rho \sigma}}^{\nu \lambda}$ being a tensor defined by \be
{g_{\rho \sigma}}^{\nu \lambda} \stackrel{\rm def.}{=}
{\delta_\rho}^\nu {\delta_\sigma}^\lambda- {\delta_\sigma}^\nu
{\delta_\rho}^\lambda. \ee The energy is expressed by the surface
integral \cite{Mo2} \be E=\lim_{r \rightarrow
\infty}\int_{r=constant} {{\cal U}_0}^{0 \alpha} n_\alpha dS, \ee
where $n_\alpha$ is the unit 3-vector normal to the surface
element ${\it dS}$.

Now we are in a position to calculate the energy associated with
 the general solution (31) using the superpotential (35).
  Thus substituting from (31) into
 (35) we obtain the following nonvanishing values
 \be
{{\cal U}_0}^{0 \alpha}={2X^\alpha \over \kappa R^3}
\left[2m\left(1-e^{-R^3/r1^3}\right)-R^2 {\cal B}' \right].
 \ee
 Substituting from (39) into
(38) we get \be E(R)=2m \left(1-e^{-R^3/r1^3}\right)-R^2 {\cal
B}'.
 \ee
 As is clear from (40), the energy content depends on the arbitrary
function ${\cal B(R)}$. If ${\cal B(R)}=1$ then the energy content
(40) will coincide with that
 of solution (17) \cite{Ga}, and if ${\cal B(R)}$  takes the form (34) then the
 energy content will coincide with that of solution (18)  \cite{Ga,Yi,Ra}.

\newsection{Concluding Remarks}

In this paper we have obtained the general solution of M\o ller's
tetrad theory of gravitation in case of spherical symmetry and
when the stress-energy momentum tensor has a specific form. The
previously obtained  solutions have been verified as a special
case of the general solution. The general solution gave no more
than the spherically symmetric nonsingular black hole solution
\cite{Ga,Di}.

The general solution has been found to contain an arbitrary
function and two constants of integration. Hence M\o ller's theory
does not fix the tetradic geometry in case of spherical symmetry;
up to finite number of arbitrary constants. It is of interest to
note that in the case of spherical symmetry when the stress-energy
momentum tensor is vanishing \cite{MWLH} the general solution does
not fix the tetradic geometry too. Also in the cases of cosmology
and stationary axisymmetric \cite{Se3,Se1} although no general
solution is obtained, it is proved that the solutions do not fix
the field equations of M\o ller's theory.  S$\acute{a}$ez in his
scalar tetradic theory of gravitation \cite{Se2} has discussed the
point that if the tetradic geometry can be fixed from the field
equations of the  Weitzenb{\rm $\ddot{o}$}ck theory. He did not
find a conclusive answer. Indeed, the present work has the
advantage that it gives a reasonable reply for his question.

The energy content of the general solution is calculated. It is
found that the energy depends on the asymptotic behavior of the
arbitrary function ${\cal B(R)}$. If ${\cal B(R)}\sim 1/R$ it will
contribute to the energy content (40) and  if ${\cal B(R)}\sim 1$
it will not contribute to the total energy (40).

The general solution of a tetrad having spherical symmetry with
three unknown functions is obtained. As for the general tetrad
having spherical symmetry which is given by Robertson \cite{Ro},
the general solution is obtained in the vacuum case \cite{SNH},
but when the stress-energy momentum tensor is not vanishing the
general solution is has not yet been obtained. This will be done
in a future research.

\bigskip
\bigskip
\centerline{\Large{\bf Acknowledgements}}

The author would like to thank Professor I.F.I. Mikhail; Ain Shams
University,  Professor M.I. Wanas, Professor M. Melek; Cairo
University; Professor T. Shirafuji; Saitama University and
Professor K. Hayashi; Kitasato University for their stimulating
discussions.

\end{document}